\documentclass[aps,prb,twocolumn,footinbib,superscriptaddress,floatfix,showkeys]{revtex4}

\usepackage{color}
\usepackage{epsfig}
\usepackage{latexsym}
\usepackage{amssymb}
\usepackage{amsmath}
\usepackage{algorithm}
\usepackage{algorithmic}
\usepackage{latexsym}
\usepackage{fancyhdr}
\usepackage{verbatim}
\usepackage{tabularx}
\usepackage{epsfig}
\usepackage{amsmath}
\usepackage{amssymb}
\usepackage{graphicx}
\usepackage{wasysym}
\usepackage{subfigure}
\usepackage{setspace}
\usepackage{xr}

\usepackage{tgheros}

\usepackage[T1]{fontenc}

\definecolor{bv}{RGB}{52,43,125}

\begin{document}

\title{Medical Implications of Space Radiation Exposure Due to Low Altitude Polar Orbits}
\author{\color{bv}{Jeffery C. Chancellor}}
\affiliation{Department of Physics and Astronomy, Texas A\&M University, College Station, Texas 77843-4242, USA}

\author{\color{bv}{Serena M. Au\~n\'on-Chancellor}}
\affiliation{University of Texas Medical Branch, Galveston, 77555, USA}
\affiliation{National Aeronautics and Space Administration (NASA), Johnson Space Center, Houston,77058, USA}

\author{\color{bv}{John B. Charles}}
\affiliation{National Aeronautics and Space Administration (NASA), Johnson Space Center, Houston,77058, USA}

\keywords{prodromal, low-Earth orbit, acute radiation, space radiation, aerospace medicine, solar particle event}

\maketitle

{\noindent \color{bv}{\bf  Introduction}: Space radiation research has progressed rapidly in recent years, but there remain large uncertainties in predicting and extrapolating biological responses to humans. Exposure to cosmic radiation and Solar Particle Events (SPEs) may pose a critical health risk to future spaceflight crews and can have a serious impact to all biomedical aspects of space exploration. The relatively minimal shielding of the cancelled 1960's Manned Orbiting Laboratory (MOL) program's space vehicle and the high inclination polar orbits would have left the crew susceptible to high exposures of cosmic radiation and high dose-rate SPEs that are mostly unpredictable in frequency and intensity.\newline
{\bf Methods}: In this study, we have modeled the nominal and off-nominal radiation environment that a MOL-like spacecraft vehicle would be exposed to during a 30-day mission using high performance, multi-core computers.\newline
{\bf Results}: Projected doses from a historically large SPE (e.g. the August 1972 solar event) have been analyzed in the context of the MOL orbit profile, providing an opportunity to study its impact to crew health and subsequent contingencies\newline
{\bf Discussion}: It is reasonable to presume that future commercial, government, and military spaceflight missions in low-Earth orbit (LEO) will have vehicles with similar shielding and orbital profiles. Studying the impact of cosmic radiation to the mission's operational integrity and the health of MOL crewmembers provides an excellent surrogate and case-study for future commercial and military spaceflight missions.}

\section*{Introduction}
\begin{figure}[hb!]
\centering
\includegraphics[width=\linewidth]{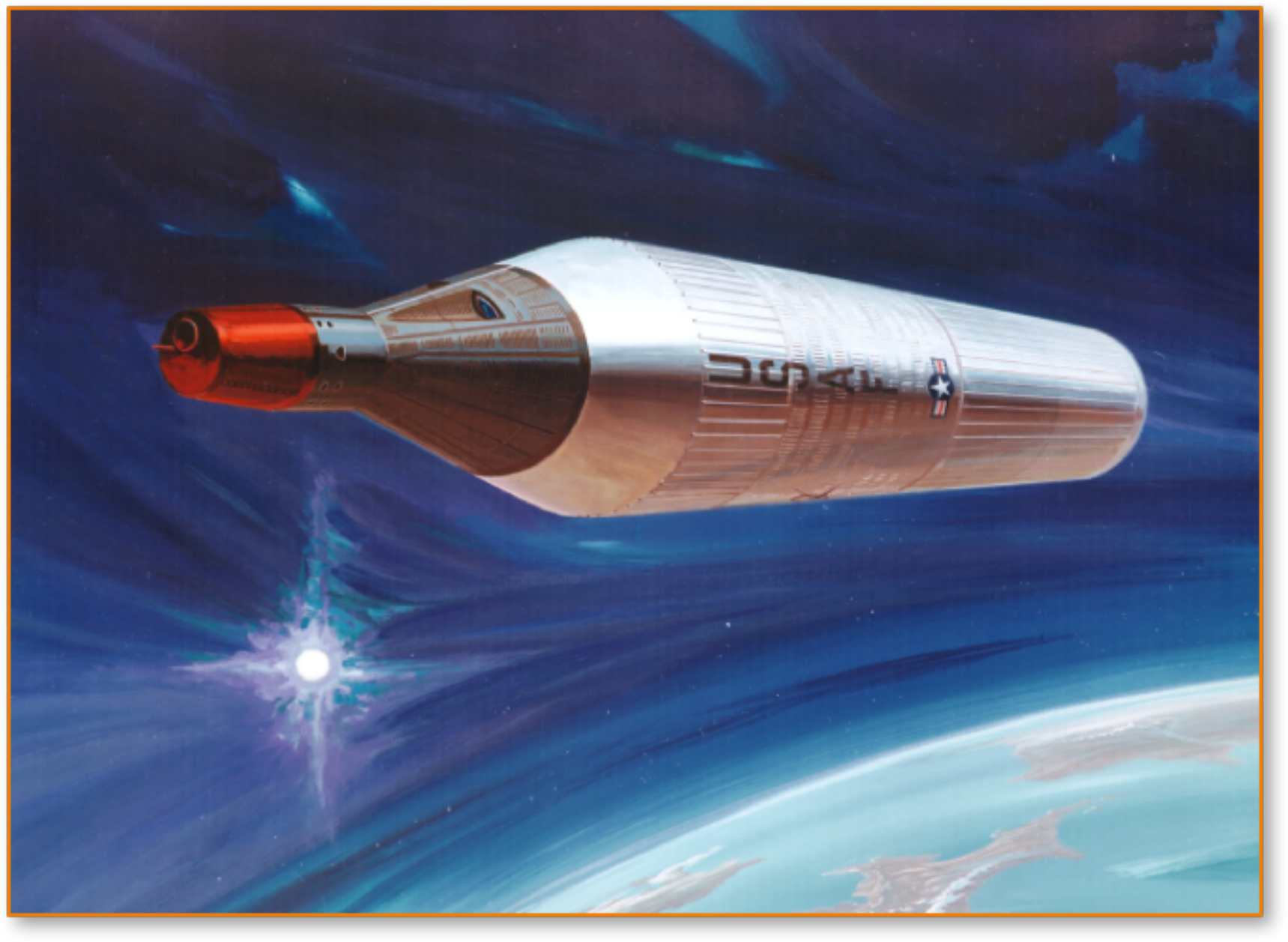}
\caption{{\bf MOL vehicle.} Artist's depiction of the proposed MOL vehicle platform (Douglas Aircraft Co., 1967).}
\label{fig:orbit}
\end{figure}

The Manned Orbiting Laboratory (MOL) program was conceived in 1963 to define and demonstrate the performance capabilities of humans in the novel environment of spaceflight and initially included many technological, observational, and biomedical investigations to establish appropriate benchmarks\cite{Charles} (\emph{e.g.} Figure \ref{fig:orbit}). Subsequently, in 1965, MOL was recast as a secret reconnaissance platform to place two military astronauts and an advanced camera system into low earth orbit for 30 days. The new mission was to demonstrate the military value of men in space through their ability to acquire high-resolution photography of America's Cold War adversaries efficiently and effectively. The program would have provided significant, reproducible and well-documented physiological and psychological stressors to its pilots permitting detailed evaluation of the effects of extended spaceflight. However, the program was cancelled in 1969 when unmanned satellites were already providing comparable data at less expense.

The MOL program saw the initiation and development of novel approaches and made lasting contributions in the areas of in-flight radiation assessment, nutritional and hygienic support, planning of workloads and rest periods with a minimum of real-time assistance from Earth, and meaningful exercise to counter the effects of extended and uninterrupted weightlessness. These contributions improved the success of operations of NASA's Apollo, Skylab, and Space Shuttle programs and continue to be utilized today aboard the International Space Station (ISS).

The radiation environment of the MOL flights has prompted specific curiosity given its unprecedented nature for manned spaceflight. Except for small amounts of radioisotopes used in manned space missions for instrument calibration and research, the vast majority of crew exposures are due to the complex radiation environment in which they must travel and live. The space radiation environment in low-Earth orbit (LEO) can be divided into three separate sources of ionizing radiation: solar wind, consisting of mostly low energy protons and electrons; heavy-charged particles found in the Galactic Cosmic Ray (GCR) spectrum; and energetic protons associated with a Solar Particle Event (SPE). The fluence (the number of incident particles crossing a given plane) of GCR particles in interplanetary space fluctuates inversely with the solar cycle, with dose-rates ranging between 50-100 mGy/year at solar maximum and 150-300 mGy/year at solar minimum\cite{Chancellor,Mewalt}. Here \emph{mGy} is the abbreviation for milligray, where the \emph{Gray} is the SI standard unit for the measurement of ionizing radiation dose (1 mGy = 0.001Gy)\cite{Hall}.The occurrence of SPEs is unpredictable but dose rates as high as 1500 mGy/hour have been measured\cite{SRAG,NCRP2006}. The background dose rate for solar protons (\emph{e.g.}, the solar wind) varies with the solar cycle (9-14 years, average 11 years per cycle), but even at solar maximum the dose rate is much less than the GCR dose rate and therefore it is considered to be of negligible risk.

 \begin{table}
\caption{The parameters that describe the simulated Manned Orbiting Laboratory mission orbit (adapted from Charles \emph{et al.}\cite{Charles})}
\vspace*{0.5em} 
\vspace*{0.5em} 
\small
\centering
\begin{tabular}{l|l}
\textbf{Orbit Parameter}	& \textbf{Value} \\
\hline
Apogee		& 344.5 km\\
Perogee		& 148.20 km\\
Inclination & 96.5$^{\circ}$\\
Altitude (highest) & 344.5 km\\
Altidude (lowest) & 148.2 km\\
Ascending Node	& 351.88$^{\circ}$\\
Argument of Perigee	& 196.77$^{\circ}$\\
True Anomaly		& 152.48$^{\circ}$\\
Eccent ricity & 0.01\\
Number of Orbits & 16.13 per day\\
Launch & July 15, 1972\\
Landing & August 14, 1972\\
\end{tabular}
\label{tab:parameters}
\end{table}

One of the most important questions to be answered for future NASA, commercial, and military spaceflight missions focuses on the short and long-term health effects of space radiation on participants. Commercial, government, and military spaceflight crews could be exposed to SPEs that might induce prodromal effects including fatigue, malaise, nausea, and vomiting, and further exacerbate biological outcomes from the concurrent chronic GCR environment. The indigenous shielding provided by the Earth's magnetic field attenuates the major effects of space radiation exposures for current NASA missions which orbit mostly below it. Additionally, the relatively low (51.6$^{\circ}$) inclination of the International Space Station provides significant protection and is responsible for this attenuation of radiation exposure in current missions. The (proposed) MOL flights intended for a polar orbit would not have had this luxury and would have been susceptible to high-energy charged particles penetrating the Earth's Magnetosphere at such latitudes. Each charged particle has the ability to damage critical cellular components when passing through the tissues of the body. In addition, neutrons produced by interactions of cosmic rays passing through the spacecraft structure can be highly penetrating and deliver a significant dose to critical organ systems. It is reasonable to presume that future commercial, government, and military spaceflight missions may have vehicles with similar shielding and polar orbital profiles, leaving the crew exposed to high fluences of cosmic radiation and high dose-rate SPEs that are unpredictable in frequency and intensity. We sought to model the radiation exposure that would have occurred during a planned MOL mission in order to understand the potential short- and long-term health effects on exposed crewmembers and to provide context for future spaceflights of similar duration and orbital parameters.

\begin{figure*}[!ht]
\centering
\includegraphics[width=\linewidth,height=10cm,keepaspectratio]{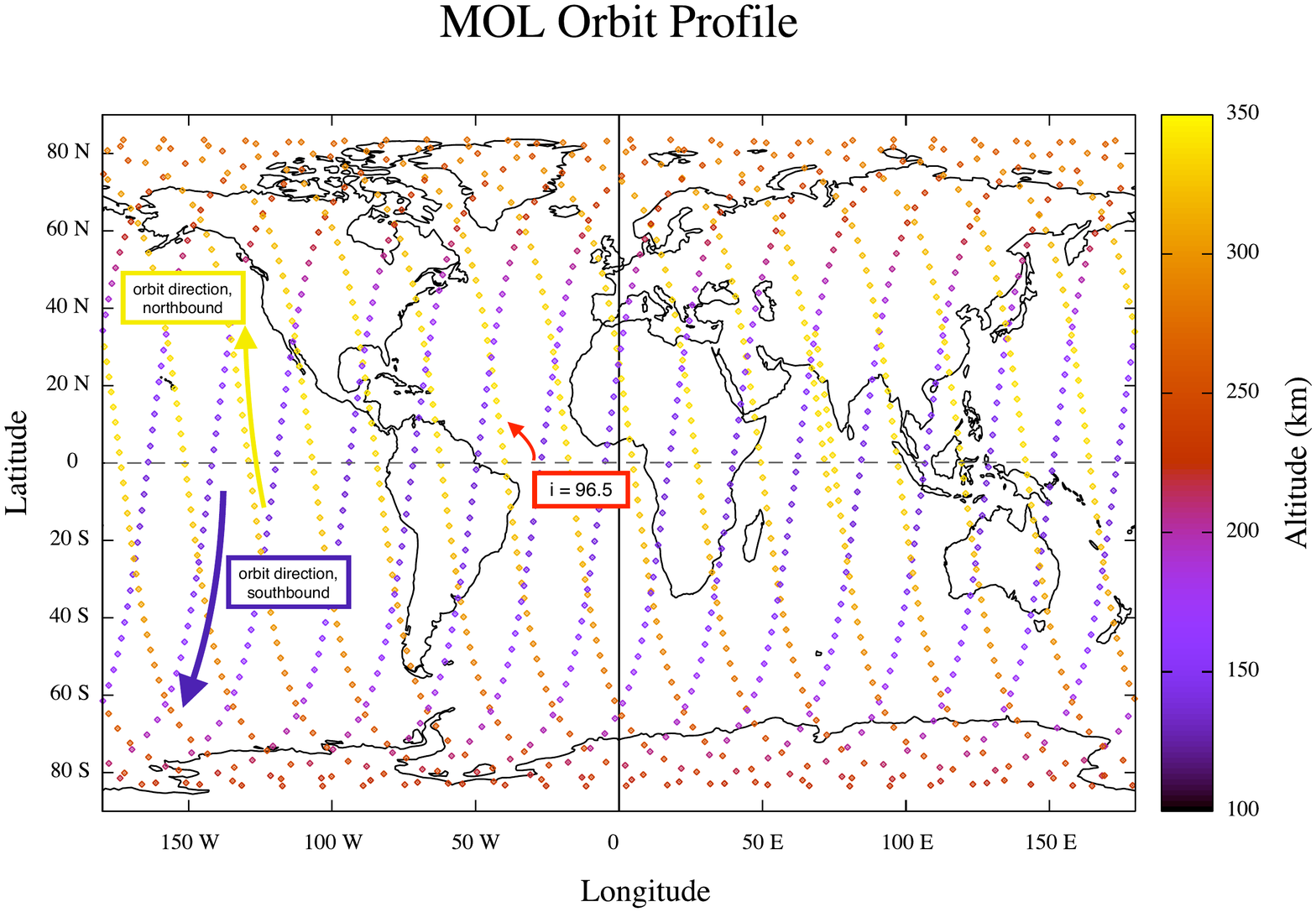}
\caption{The MOL mission profile as a function of orbit longitude, latitude and altitude. The high inclination orbit requires passes directly over the northern and southern polar regions. The inclination, shown on the figure as \emph{i}, is defined as the angle between the orbital path and the Earth's equator. The large arrows highlight the north versus southbound direction of the ground tracks. Here we can easily see that lower altitudes correspond to the area around Russia and the highest altitudes are during transversal of the polar regions, minimizing the exposure to cosmic rays and energetic solar protons.}
\label{fig:groundtracks}
\end{figure*}

\section{Methods}
Advanced numerical methods and high-performance computing capabilities allow for an accurate simulation of multiple environmental factors. These include the orbital path as a function of longitude, latitude, and altitude; the geomagnetic profile and field strength along the orbit; fluctuations in the density of space radiation due to geomagnetic field strength; and variations in solar activity. We have derived these parameters using widely-accepted standard models: the IGRF-12 geomagnetic field model\cite{thebault2015international}, the AE9/AP9 models that describe Earth's trapped radiation environment\cite{ginet2012new}, the King and CREME96 solar proton models for nominal and contingency solar proton flux\cite{king1974solar,CREME}, the ISO 15390 Standard Model for GCR flux\cite{nymmik1996galactic}, and PHITS\cite{Niita2006} (Particle and Heavy Ion Transport System) for approximating the dose behind shielding material. The details of these models can be found in the referenced text and will not be discussed in this report.

The mission ground tracks were reproduced with the sun-synchronous orbit profile shown in Table \ref{tab:parameters} resulting in the approximate trajectory shown in Figure \ref{fig:groundtracks}. This high inclination orbit (96.5$^{\circ}$) is commonly referred to as a \emph{polar orbit}. This provided the necessary input for the IGRF-12 model to determine geomagnetic profile and field strength along the orbit path. This determined the particle cutoff threshold and attenuation for the trapped protons, SPE protons, and GCR nuclei that would compromise the local radiation field.

\begin{figure*}[!hbt]
 \begin{center}
   \subfigure[Proton fluences for both test cases used.]{\label{fig:figure3}\includegraphics[width=.4\textwidth,height=4.7cm]{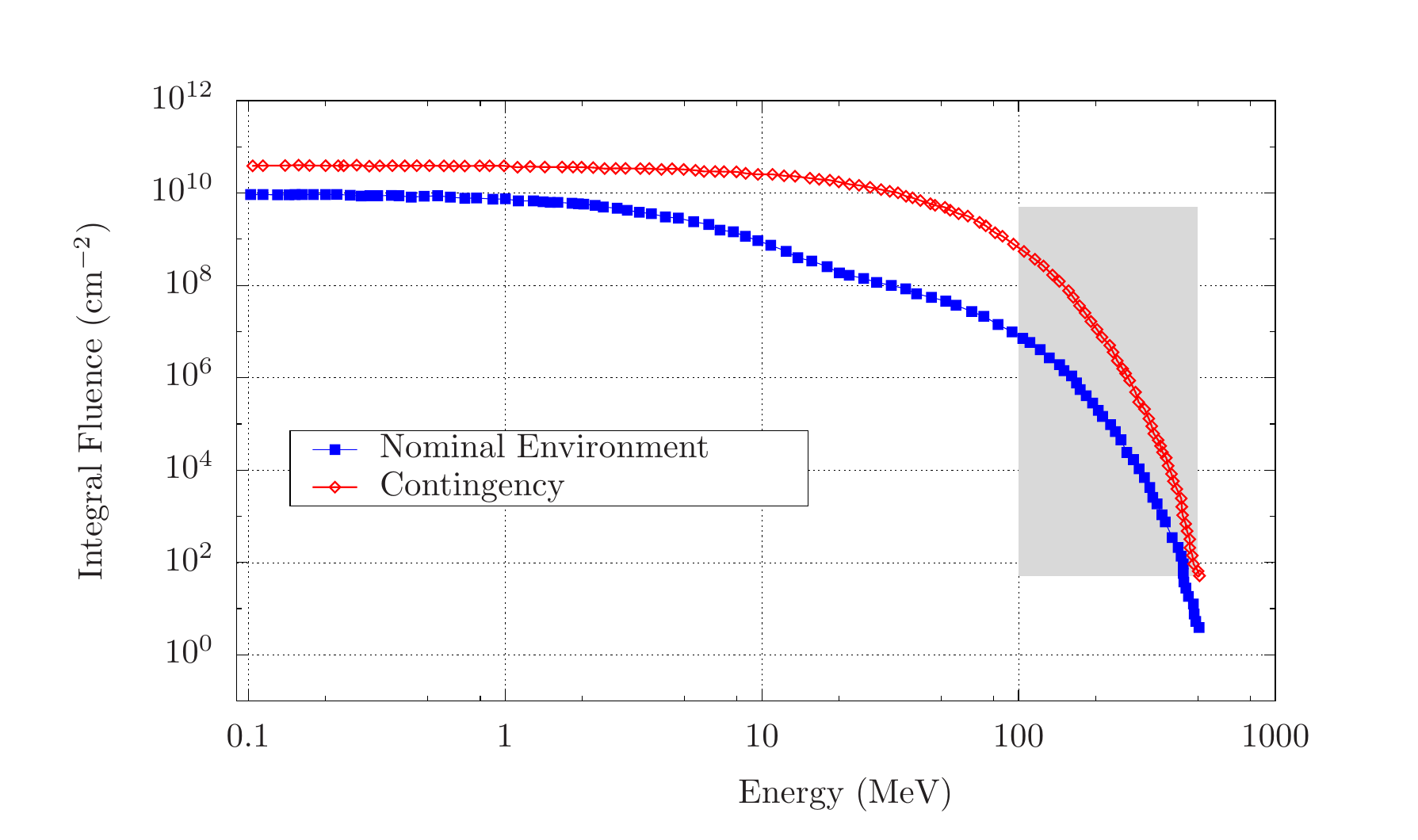}}
   \subfigure[The resulting IVA (skin) dose-equivalent for both cases.]{\label{fig:figure4}\includegraphics[width=.4\textwidth]{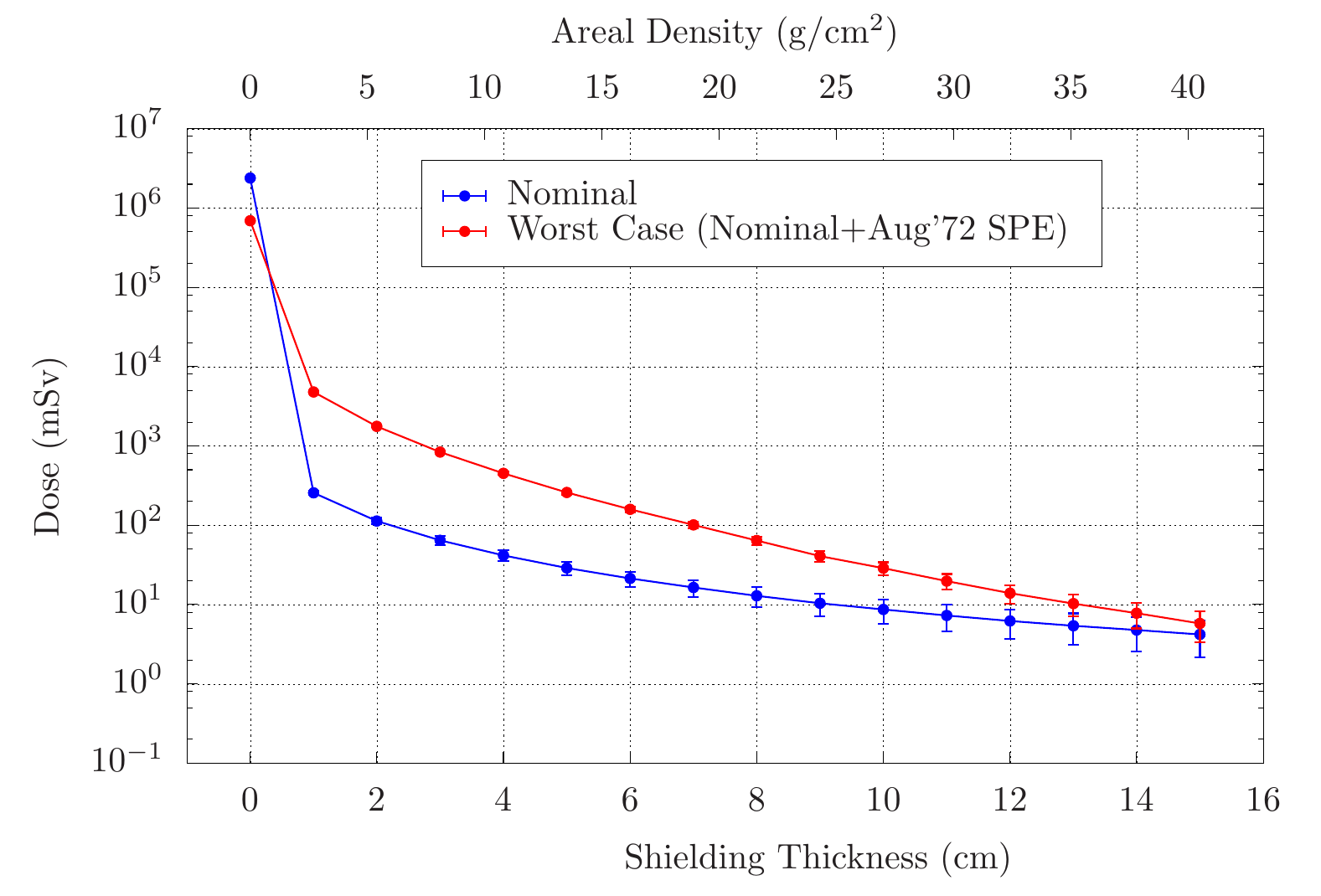}}
 \end{center}
 \caption{(a) Proton fluences for both the nominal and contingency test cases. The gray shaded box emphasizes the enhancement of $\geq$ 100 MeV protons for the case where the August 1972 SPE is included in the dose projections. The $\geq$ 100 MeV protons can penetrate typical spacecraft shielding with enough energy to reach bone marrow and BFO depths in tissue. (b)
 The resulting dose-equivalent as a function of shielding thickness and areal density. The MOL vehicle had somewhere between 5 g/cm$^{2}$ to about 15 g/cm$^{2}$ (or $\approx$ 2--6 cm) of shielding, with the lower shielding the most likely configuration. The graph shows that the addition of the August 1972 SPE increases the mission dose by close to one order of magnitude. Note that there is minimal difference for areal densities $\geq$ 30 g/cm$^{2}$, the approximate average shielding of the massive International Space Station.}
 \label{fig:radiation}
\end{figure*}

Two test cases were performed for typical and worst-case mission scenarios (hereafter referred to as nominal and contingency, respectively). The nominal test case accounted for the radiation spectrum impinging the MOL vehicle along its orbit over the course of its 30-day mission. The radiation field included GCR nuclei, trapped protons and electrons, solar wind nuclei, and one small size SPE. The small size SPE was added for accuracy since it was statistically likely that a small- to medium-sized SPE event would occur over the planned mission duration. The contingency test case included the radiation field from the nominal test case and added the proton contribution from the infamous and large SPE that hit the LEO environment on August 4, 1972, described by the spectrum,

\begin{equation}
  J(>E) = 7.9x10^{9}\operatorname{e}^{(30-E)/26.5}
\end{equation}
for protons with energies between 10-200 MeV (mega-electron volt). The electronvolt is the energy gained by an electron accelerated through a potential difference of 1 Volt\cite{Hall}. In SI units, the electronvolt is equivalent to approximately 1.602176 x 10$^{-19}$ Joules (1 MeV = 1,000,000 eV). Here the fluence $J$ is given in cm$^{-2}$ and the energy, $E$, in MeV\cite{king1974solar}. The resulting radiation fields for the nominal and contingency test cases were both integrated over the duration of the MOL mission and then the vehicle shielding was applied using the Monte Carlo-based particle transport platform, PHITS, to determine the intravehicular dose. The integrated proton fluences for both cases are shown in Figure \ref{fig:figure3}.

The August 4, 1972, SPE event is interesting for space radiation studies because the proton spectrum includes a large contribution from protons with energies exceeding 100 MeV. A 100 MeV proton has sufficient energy to penetrate typical spacecraft shielding (5-10 g/cm$^{2}$) and still have enough remaining energy to reach bone marrow and blood forming organ (BFO) depths. In fact, this SPE accounted for approximately 83\% of the $\geq$ 100 MeV protons measured during solar cycle 20, which lasted from approximately October 1964 to March 1976\cite{king1974solar}. The August 1972 event is also relevant because it occurred during the period in which MOL missions were projected to occur. Thus, our consideration of its implications reflects a reasonably probable event for the MOL program if it had been implemented. The physiological, behavioral, and operational results of the MOL program have been discussed by Jenne\cite{Jenne}.

The true value of MOL vehicle shielding is currently not available and the recently declassified and available program documents do not sufficiently describe the vehicle material and thickness. For this study, we assumed the MOL vehicle was similar to the Skylab vehicles-which almost certainly overestimated MOL shielding capacity-approximated an isotropic (e.g. the same value in all directions) shielding of 5 g/cm2. For perspective, the Apollo crew vehicle shielding was 5-10 g/cm2, while the ISS is approximately 30-50 g/cm2 of shielding mass\cite{Walker}.

The bootstrap method was utilized for error analysis to verify the statistical stability of the results and minimize systematic biases in the outcomes\cite{athreya1987}. Additionally, some validation of our results was done by applying our methods to the orbital profile of the Skylab missions (with its much lower 50$^{\circ}$ inclination). The recorded dose for the Skylab 4 mission was 178 mSv (milli-Sievert) for an 83 day mission\cite{SRAG}. Our model approximation determined a mission dose of 152 mSv, or within 15\% of the actual measured dose.

The Sievert is the SI standard of measurement for equivalent dose (1 mSv = 0.001 Sv). The equivalent dose is derived from multiplying the ionizing dose (in Gray) by a weighting factor (w$_{R}$) that accounts for variations in observed outcomes of different radiations. Sv = $w_{R}$ x Gy. The weighting factor is specific to the radiation species and biologic endpoint and discussed in detail by Hall and Giaccia\cite{Hall}. All results from this study have been reported in units of millisievert (mSv) for consistency, and also for comparison with familiar clinical diagnostic and radiotherapy exposures. 

\section{Results}
The results for both the nominal and contingency test cases are shown in Figure \ref{fig:figure4} where we have determined the skin dose-equivalent for the intravehicular radiation field. For the nominal test case, the MOL crew would have received a skin dose of 113.6 mSv and an approximate BFO dose of 41.6 mSv. In the contingency scenario, our results indicate the crew would have received an exposure of 1,770 mSv and 451 mSv to the skin and BFO, respectively. 

The recent one-year mission completed by a NASA astronaut and a Russian cosmonaut occurred during a similar portion of the solar cycle as the MOL missions; this provides us with a reasonable comparison of the nominal mission exposure calculations. For the one-year mission, the dose for any 30-day period would be 17 mSv during solar maximum and 25 mSv during solar minimum\cite{Chancellor}.  In comparison, the MOL crewmembers would have received a skin dose approximately 4--5 times higher than any 30-day dose received by the one-year crew.

\section{Discusion}
Evaluation of how the radiation exposures might have affected the crew is much more difficult to perform and there are no clear clinical interpretations for either a modest or an extreme scenario similar to our contingency test-case. Current medical standards are largely based on epidemiology studies of human populations exposed to whole body irradiation at high doses and high dose-rates limited to scenarios not found during spaceflight missions. The research challenge posed to radiation researchers is outside of the scope of this study and detailed in numerous other publications. It suffices to say that research focused on space radiation induced human health effects, unlike bone health, nutrition, cognitive functions, etc., does not have a human model exposed to the environment for properly evaluating the risk, let alone clinical mitigation.

It should be noted that the medical spaceflight standards are derived implicitly for NASA astronauts who have met a rigorous standard of health. Caution should be taken when evaluating the nominal and contingency doses for non-astronaut spaceflight passengers (\emph{e.g.} commercial spaceflight tourism). Even so, some clinical outcomes can be anticipated and elucidated in the context of NASA's spaceflight health standards for preserving astronaut crew health\cite{williamsnasa}.

The contingency scenario mission dose would surpass both the 30-day and the annual limit for BFO established by NASA for radiation exposure in LEO (as seen in Table \ref{tab:limits}).  More than 90\% of the dose incurred in this scenario is due to an acute exposure to energetic protons at an average dose rate of 23 mSv/hr. This is two orders of magnitude higher than the nominal dose rate of 156.9 $\mu$Sv/hr (0.1569 mSv/hr) or the approximate average nominal dose rate of 29 $\mu$Sv/hr for the recent one-year mission. These doses are likely to induce prodromal symptoms, but not expected to be implicitly life threatening with prompt instigation of medical countermeasures. It is important to note that this conclusion is made based on the robust health requirements of current NASA astronauts and would need to be re-evaluated for individuals that do not meet those standards. 

\begin{table}
\caption{NASA 30 day, annual, and career exposure limits for astronauts compared with predicted Manned Orbiting Laboratory nominal and contingency exposures (in mSv) \cite{SRAG,williamsnasa}.}
\vspace*{0.5em} 
\small 
\centering
\begin{tabular}{m{5em}lccr}
\textbf {Program} & \textbf {Exposure Period} & \textbf{Skin} & \textbf{Eye} & \textbf{BFO}\\
\hline
 & & & &\\
{\bf NASA} & 30 Days & 1,500 & 1,000 & 250\\
 & Annual		& 3,000 & 2,000 & 500\\
 & Career & 6,000 & 4,000 & n/a \\
 & & & &\\
{\bf MOL} & 30 Days (nominal) & 113.6 & n/a & 41.6\\
 & 30 Days (contigency)& 1,770 & n/a & 451\\
\end{tabular}
\label{tab:limits}
\end{table}

The prodromal phase of acute radiation syndrome includes clinical symptoms of nausea, vomiting, and anorexia, and may manifest within 48 hours following the SPE exposure. These symptoms may also develop within a few hours of radiation exposure\cite{fajardo2001radiation}; higher SPE doses  can result in increased severity, quicker onset, and longer duration of the symptoms\cite{anno1protracted}. Emesis, fatigue, and other expected symptoms could seriously impair crew performance and mission success. Recent research results from an SPE-like proton distribution on a ferret model indicated that emesis responses were observed in doses as low as 400-1,000 mSv\cite{sanzari2013effects}. Prodromal vomiting in \emph{humans} is expected at doses greater than approximately 750-1000 mSv and is the most likely acute effect that can impact crew health after exposure to a significantly large SPE dose. For comparison, the LD50 (following an acute radiation exposure) for ferrets was determined by Harding ($<$2 Gy) and for humans by Hall and Giaccia (3--4 Gy)\cite{harding_prodromal_1988,Hall}. For SPE protons, this would translate to an approximate equivalent dose of 3.5 Sv and 4.5--6 Sv respectively. A minimal increase in fatigue in the form of depressive or anxious behaviors could manifest after radiation exposure; however, it would be highly unlikely that the doses modeled here would exacerbate fatigue or other adverse behaviors over and above baseline levels\cite{york2012biobehavioral,york2012individually,york}.

Although these doses are not by themselves expected to result in death, it is conceivable that the acute exposure to SPE protons along with other spaceflight stressors such as microgravity could exacerbate radiation-induce immune suppression, and thus could ultimately result in severe outcomes if not treated appropriately. Studies of the synergistic effects of radiation combined with spaceflight environment stressors (\emph{e.g.} microgravity, environment toxicity, emotional stress, etc.) show increased susceptibility to infection, delayed wound healing and decreased survival\cite{maks2011analysis,sanzari2011combined,wilson2012comparison}. Overall suppression of the immune system may lead to a compromise in crew health status, so that an SPE-like exposure in combination with spaceflight environment stressors could enhance the risk of pathogenic infection. Outcomes resulting from the alterations in levels of immune activation found during spaceflight, interacting with SPE-like radiation exposure(s) and subsequent immune alterations, should be evaluated with respect to other physiologic systems, including bone, muscle, endocrine, neurologic, respiratory, etc.

There are currently few medical countermeasures available for the management of the various acute injuries that could occur during spaceflight. Burn care, wound closure and treatment, management of traumatic injury, antiemetic, and infection control capabilities would likely be available, but the capability to treat multiple affected crewmembers for extended periods could quickly outstrip available medical resources. MOL planning would have permitted the crewmembers to evacuate the laboratory and return to Earth in short order. Three well-supported low-latitude recovery zones around the globe would have accommodated a daylight landing from the low polar orbit within 12 hours of the decision to terminate the mission and within six hours if a nighttime splashdown was permitted. This is comparable to the options for Earth-return currently available to ISS crewmembers\cite{anonymous}.

There are still uncertainties in the mechanisms behind the synergistic lethality observed with radiation injury and the efficacy of treatments against damage resulting from radiation-combined injury from other sources (\emph{e.g.} microgravity, infection, etc.). Presently, only limited testing has been done on the efficacy of treatment regimens on traumatic or acute injury when radiation exposure is a factor, especially charged particle radiation such as that found in the space environment. In short, the lack of human exposures to extreme doses and dose-rates of charged particle radiation limits the ability to provide sound, clinical-based interpretation of radiation-induced health effects, particularly when concomitant injury from other sources is present.

\section{Conclusion}
We have shown that the unique nature of the low Earth orbit of the Manned Orbiting Laboratory flights planned, but not flown, in the 1960's would have exposed its crews to a problematical radiation dose. While the nominal 30-day mission's exposure would have been well within current NASA limits, the contingency scenario including the August 1972 SPE would have exceed NASA's 30-day limit for skin exposure and would have been nearly double the limit to BFO. The contingency scenario may have caused transient illness in healthy and highly-conditioned MOL pilots, but similar exposures to individuals with less-than-astronaut fitness would probably cause greater distress than could be accommodated in future NASA, commercial and military LEO vehicles. Appropriate attention to mission and vehicle design and available radiation countermeasures is advisable.

\vspace*{0.5em} \noindent {\color{bv} \bf Acknowledgments}

\noindent {\scriptsize \begin{spacing}{1.00}
\noindent J.C.C. would like to thank BRCC for support and motivation during the course of this study. J.C.C. and S.M.A would also like to thank Dr. Nicholas Stoffle for his review of the manuscript and many fruitful discussions on advanced \emph{GEEH} simulations. The authors acknowledge the Texas Advanced Computing Center (TACC) at The University of Texas at Austin for providing HPC resources that have contributed to the research results reported within this paper.

Part of the work of J.C.C.~ has based upon work supported by the Office of the Director of National Intelligence (ODNI), Intelligence Advanced Research Projects Activity (IARPA), via Interagency Umbrella Agreement IA1-1198. The views and conclusions contained herein are those of the authors and should not be interpreted as necessarily representing the official policies or endorsements, either expressed or implied, of the ODNI, IARPA, or the U.S. Government. The U.S. Government is authorized to reproduce and distribute reprints for Governmental purposes notwithstanding any copyright annotation thereon.
\end{spacing}}

\vspace*{0.5em} \noindent {\color{bv} \bf Author Contributions} \noindent {\scriptsize \begin{spacing}{1.00}
\noindent All authors contributed to the content and review of this manuscript.

\end{spacing}}

\vspace*{0.5em} \noindent {\color{bv} \bf Competing financial interests} \noindent {\scriptsize \begin{spacing}{1.00}
\noindent The authors declare that they have no competing financial interests.

\end{spacing}

\noindent {}

\vspace*{-1.5em}

\bibliographystyle{unsrt}
  
\end{document}